\documentclass[iop]{emulateapj}

\usepackage[colorlinks=true,citecolor=blue,linkcolor=blue]{hyperref}

\shorttitle{UDGs in group environments}
\shortauthors{Merritt et al.}

\begin{document}


\title{The Dragonfly Nearby Galaxies Survey. II. Ultra diffuse galaxies near the elliptical galaxy NGC 5485}


\author{Allison Merritt\altaffilmark{1}, Pieter van
  Dokkum\altaffilmark{1}, Shany Danieli\altaffilmark{1,5,6}, Roberto
  Abraham\altaffilmark{2,3}, Jielai Zhang\altaffilmark{2,3},
  I. D. Karachentsev\altaffilmark{7}, L. N. Makarova\altaffilmark{7}}


\altaffiltext{1}
{Department of Astronomy, Yale University, 260 Whitney Avenue, New Haven, CT 06511, USA}
\altaffiltext{2}
{Department of Astronomy and Astrophysics, University of Toronto,
50 St. George Street, Toronto, ON, Canada M5S~3H4}
\altaffiltext{3}
{Dunlap Institute for Astronomy and Astrophysics, University of Toronto, Toronto ON,
M5S 3H4, Canada}
\altaffiltext{4}
{Canadian Institute for Theoretical Astrophysics, Toronto, ON, M5S 3H4, Canada}
\altaffiltext{5}
{Yale Center for Astronomy and Astrophysics, Yale University, New Haven, CT 06520, USA}
\altaffiltext{6}
{Department of Physics, Yale University, New Haven, CT 06520, USA}
\altaffiltext{7}
{Special Astrophysical Observatory, Nizhnij, Arkhyz, Karachai-Cherkessia 369167, Russia}


\begin{abstract}
We present the unexpected discovery of four ultra diffuse galaxies (UDGs) in a group environment. We recently identified seven extremely low surface brightness
galaxies in the vicinity of the spiral galaxy M101, using data from the
Dragonfly Telephoto Array. The galaxies have effective radii of $10''-38''$ and central
surface brightnesses of $25.6-27.7$ mag arcsec$^{-2}$ in g-band. We subsequently
obtained follow-up observations with $HST$ to constrain the distances to these galaxies.
Four remain persistently unresolved even with the spatial resolution of $HST$/ACS, 
which implies distances of $D > 17.5$ Mpc. We show that the galaxies are most likely
associated with a background group at $\sim 27$ Mpc containing the massive ellipticals
NGC 5485 and NGC 5473. At this distance, the galaxies have sizes of $2.6-4.9$ kpc,
and are classified as UDGs, similar to the populations that have
been revealed in clusters such as Coma, Virgo and Fornax, yet even more diffuse.
The discovery of four UDGs in a galaxy group demonstrates that the UDG phenomenon is not
exclusive to cluster environments. Furthermore, their morphologies seem less
regular than those of the cluster populations, which may suggest a different formation
mechanism or be indicative of a threshold in surface density below which UDGs are unable
to maintain stability.
\end{abstract}

\section{Introduction}
The lowest detectable surface brightnesses of galaxies are, in
practical terms, a function of survey depth
\citep{Disney1976,Dalcanton1995}. Low surface brightness galaxies
(LSBGs) are known to exist at all sizes
\citep[e.g.][]{Zucker2006,McGaugh1994,Bothun1987} and across all 
environments, from the field \citep{Impey1996} to the Local Group
\citep{McConnachie2012} to massive clusters such as Virgo and Coma
\citep[][]{Ulmer1996,Impey1988,Caldwell2006,Adami2006,Davies2015}. When
compared to high surface brightness galaxies (HSBGs) of either
similar luminosity or effective radius, the integrated number density
of LSBGs surpasses that of HSBGs \citep{Dalcanton1997}. And, as
advances in instrumentation \citep[e.g.][]{Abraham2014} and
methodology allow observations to push down to ever lower surface
brightness limits, the diversity of the low surface brightness
universe is continuously unveiled. 

Recently, a population of ultra diffuse galaxies (UDGs) featuring
extremely low central surface brightnesses ($\mu_{g,0} > 24$ mag
arcsec$^{-2}$) and large effective radii ($R_{e} > 1.5$ kpc) was
identified in the outskirts of the Coma cluster by \cite{vanDokkum2015a}; and
subsequent searches in the Virgo \citep{Mihos2015} and Fornax
\citep{Munoz2015} clusters revealed similar (though less numerous)
populations. 

\begin{figure*}[!t]
\begin{center}
\includegraphics[width=\textwidth]{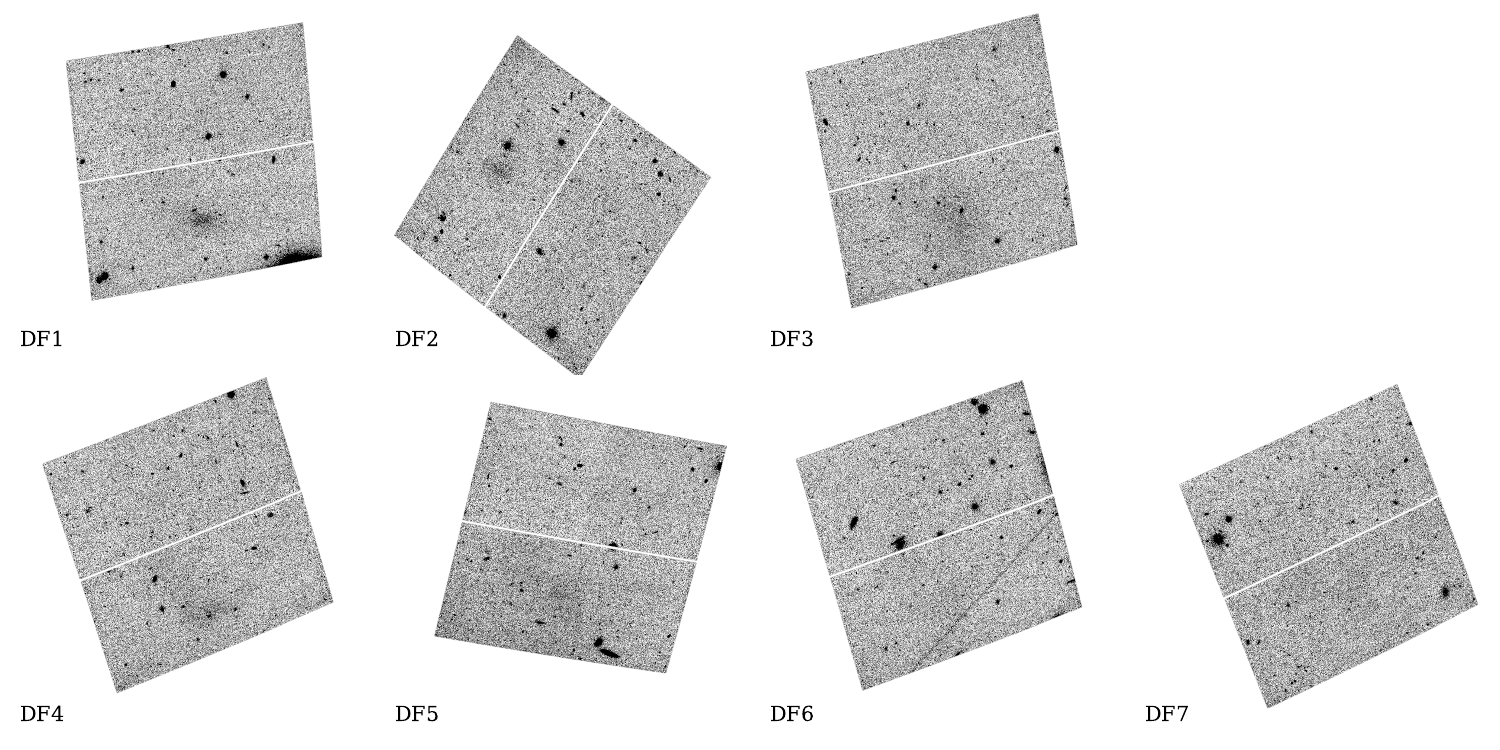}
\caption{HST images of all seven LSB galaxies reported in
  \cite{Merritt2014}. Observations were obtained in F606W and F814W,
  with 0.5 orbits each, although here we show only F606W. North is up and East is to the
  left. The field of view is $\sim 3.5$ arcminutes on a side.
\label{hstfig}}
\end{center}
\end{figure*}

\cite{vanDokkum2016} measured the stellar velocity dispersion of Dragonfly 44
(Coma-DF44,hereafter DF44), one of the largest UDGs in the Coma cluster, and showed that the
dark matter fraction within the effective radius is $98$\%. Similarly, globular cluster
counts in two other UDGs $-$ Dragonfly 17 (Coma-DF17, hereafter DF17) in Coma
\citep{PengLim2016,BeasleyTrujillo2016} and VCC 1287 in Virgo
\citep{Beasley2016} $-$ indicate that these are systems with unusually high
total mass-to-light (or total-to-stellar mass) ratios for their masses. While it seems
clear that all three of these galaxies are underluminous for their total mass (or
globular cluster count), UDGs do \textit{not} all have the same mass. DF44 is likely hosted
by a massive ($\sim 10^{12} M_{\odot}$) dark matter halo, whereas the other two UDG host
halos are less massive, with reported estimates between
$8\times10^{10}M_{\odot}-10^{11}M_{\odot}$ \citep[see][]{vanDokkum2016}.

It is not yet understood how UDGs form, but considering that UDGs are empirically defined
solely on the basis of their size and surface brightness, it is reasonable to expect that
multiple formation mechanisms may be at play. The UDGs VCC 1287, Dragonfly 44, and
Dragonfly 17 seem to be ``failed'' galaxies that are underluminous for their mass
\citep{BeasleyTrujillo2016,vanDokkum2016}. There could plausibly be additional,
fundamentally different systems that fall into the UDG category as well.
\cite{AmoriscoLoeb2016} suggest, for example, that dwarf galaxies residing in dark matter
halos with high spin have large sizes (and thus low surface brightnesses) and would therefore be observationally classified as UDGs. Any such objects are likely distinct from DF44, DF17 and VCC 1287, however, as all three of these galaxies are large and underluminous relative to their globular cluster systems.
Finally, some UDGs could also be  normal dwarfs disrupting in harsh cluster environments \citep{Moore1996}, analogous to disrupting dSphs in the Local Group \citep[e.g.][]{Collins2013}.

In addition to mass measurements, understanding UDGs and their properties as a
function of environment is key to determining how and where UDGs form. This is a nontrivial
task, however, as the low surface brightness nature of UDGs means that obtaining reliable
distance measurements ranges from difficult to nearly impossible. \cite{Dalcanton1997}
obtained spectroscopic redshifts of seven
field LSBGs, two of which are large enough and faint enough to qualify as UDGs. More recently,
\cite{Toloba2016} and \cite{Crnojevic2016} measured the distances to relatively nearby
(and likely disrupting) UDGs associated with the spiral galaxy NGC 253 and the massive 
elliptical NGC 5128 (Cen A), respectively, via the Tip of the Red Giant Branch method. 
Finally, \cite{Makarov2015} and \cite{MartinezDelgado2016} were able to spectroscopically
confirm the presence of UDGs in low density regions.

In this paper, we present evidence for the existence of UDGs in a group
environment. We use $HST$ observations to constrain the distances to seven previously
identified LSBGs \citep{Merritt2014}, and find that four of the seven must lie at
distances $>17.5$ Mpc. The lower limits on distance translate into minimum sizes, and we
classify these galaxies as UDGs. We assess the likely environment
of the sample and discuss the implications of a population of group UDGs.

\begin{figure*}[!t]
\begin{center}
\includegraphics[width=\textwidth]{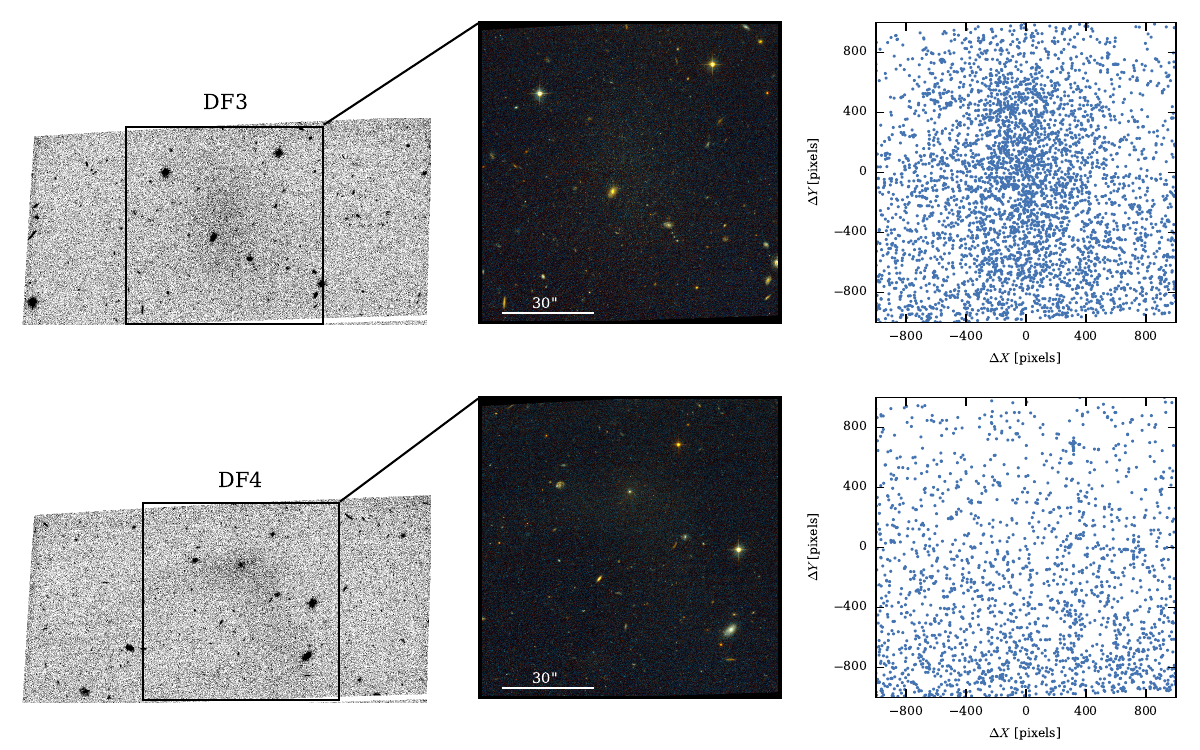}
\caption{Left: greyscale $HST$ images of DF3 and DF4 ($F606W$). These two LSBGs have very similar sizes and central surface brightnesses (see Table \ref{paramstable}). 
Middle: zoomed in color images of DF3 and DF4, created from $F606W$ and $F814W$ images, show that the galaxies are very different at $HST$ resolution $-$ DF3 resolves into a sea of faint stars, whereas DF4 appears to be an empty image. This is demonstrated more clearly in the far right panel, where we show the distribution of point sources in both images as determined by DOLPHOT. \label{ptsourcefig}}
\end{center}
\end{figure*}

\section{Imaging} 

\subsection{Dragonfly}
The diffuse galaxies examined in this work were originally discovered
in data taken by the Dragonfly Telephoto Array \citep{Abraham2014}, in
a field centered on the nearby spiral galaxy M101 (\citealt{Merritt2014};
we note that the galaxies are also visible in images taken on $0.1-0.8$ m amateur
telescopes by \citealt{Karachentsev2015} and \citealt{Javanmardi2016}).
Dragonfly is a robotic, refracting telescope
designed specifically for the detection of extended, extremely low
surface brightness optical emission.

The raw data frames were taken in the Spring of 2013, for a total of
$\sim35$ hours. Full details of data collection and the data reduction
pipeline are given in \cite{vanDokkum2014} and \cite{Merritt2014}. In
brief, we obtain calibration frames each night and apply these to the
individual frames. We model and correct for a sky gradient produced by
changes in the sky background with zenith distance with a second order
polynomial after aggressively masking all objects in the frame. 

The final reduced images are constructed from $g$-band and $r$-band
frames, and have a limiting surface brightness 
of $\mu_{g}\sim 29.5$ mag arcsec$^{-2}$ and $\mu_{r}\sim
29.8$ mag arcsec$^{-2}$ when measured in $10''$ boxes.
Star-subtracted images were also required for a robust detection of low surface
brightness galaxies \citep{Merritt2014}; these were produced using a custom pipeline that
builds and applies an empirical, spatially varying composite PSF
\citep[described in detail in][]{Merritt2016}.

\begin{figure*}[!t]
\begin{center}
\includegraphics[width=\textwidth]{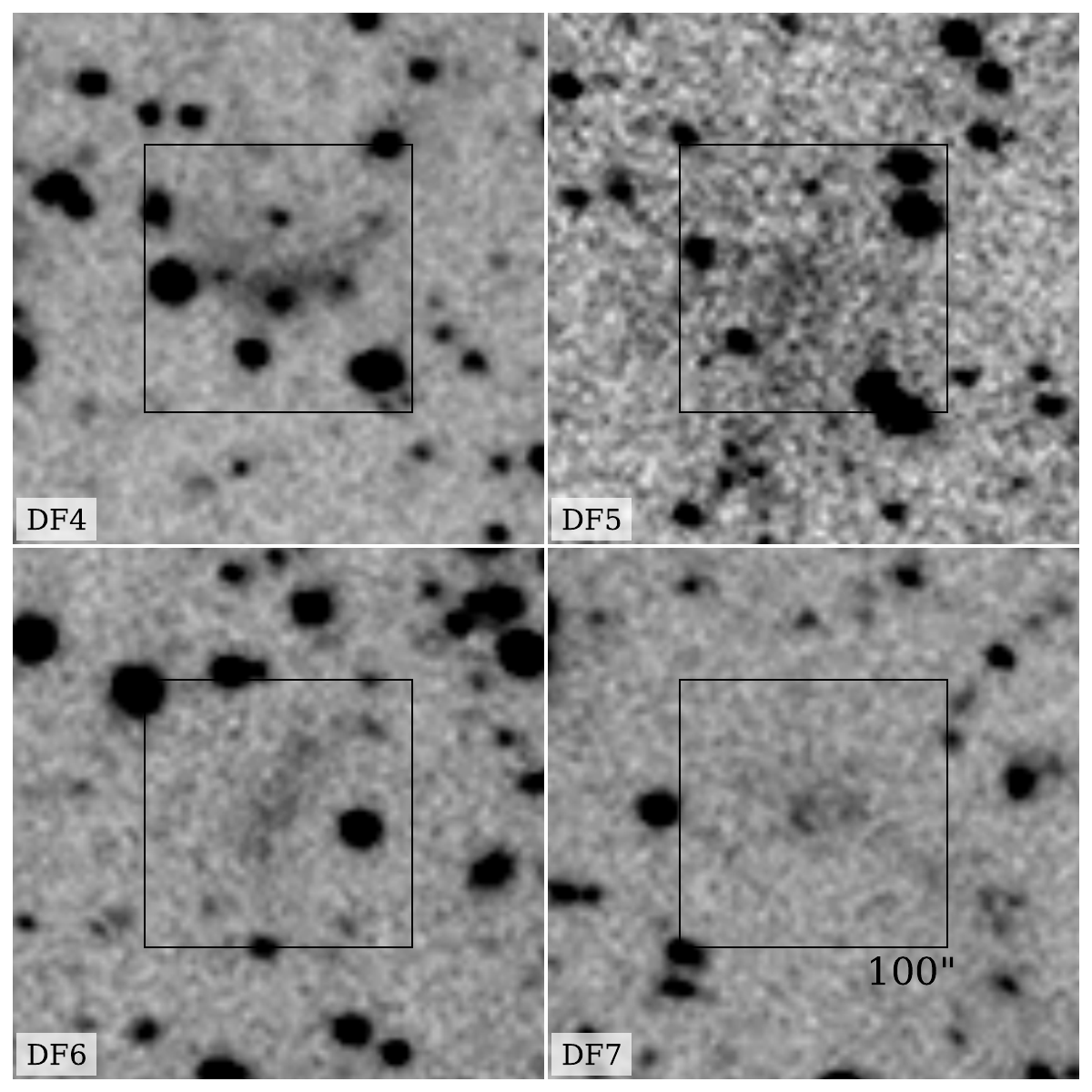}
\caption{Dragonfly discovery images of the four unresolved LSBGs. The cutouts are
$200$ arcsec on a side, with black boxes highlighting the central 100 arcsec. The
galaxies are all extremely large and diffuse, but beyond that show a high degree
of morphological diversity. 
\label{cutoutsdf}}
\end{center}
\end{figure*}

\subsection{HST}
\begin{figure*}[!t]
\begin{center}
\includegraphics[width=\textwidth]{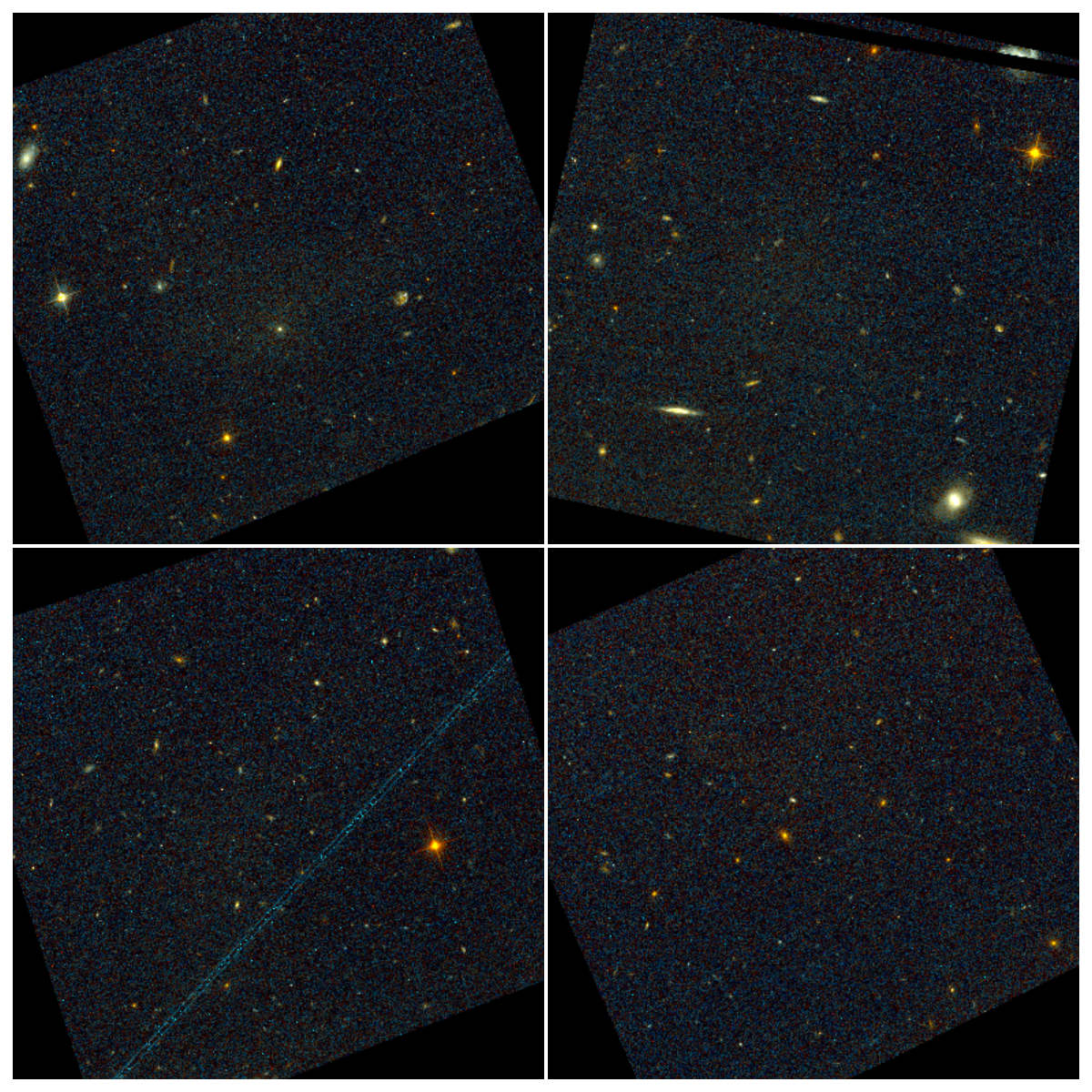}
\caption{HST pseudo-color images of the galaxies (the area corresponds to the $100$
arcsec/side boxed regions in Figure \ref{cutoutsdf}), created from the F606W and F814W
images. The four unresolved galaxies are barely visible in these data, although
they re-emerge in binned data (Figure \ref{datafig}).
\label{cutoutshst}}
\end{center}
\end{figure*}
The optical colors, luminosities, and morphologies of the
seven LSBGs are consistent with being satellite galaxies associated
with M101 itself; and, as described in \cite{Merritt2014}, none possess the
large-scale wispy structure characteristic of galactic cirrus
\citep[e.g.][]{Guhathakurta1989}. We cannot derive
their distances from Dragonfly photometry alone, however.
We therefore obtained follow-up data with $HST$, as the higher spatial
resolution ($0.05$ arcsec pixel$^{-1}$) should provide photometry of resolved stars
in each of these galaxies if they are in fact members of the M101 group
\citep[at a distance modulus of $29.04$, or $\sim 6.4$ Mpc;][]{Shappee2011}. 

The $HST$ data were obtained through the $HST$ program 13682 and
consist of ACS/WFC imaging. Each galaxy was observed for $0.5$ orbits
with both the $F606W$ and $F814W$ filters. The data were reduced with the
default $HST$ reduction pipeline, and the drizzled $HST$ images (generated with
\texttt{Astrodrizzle} from the indivual calibrated images) for the seven galaxies
are shown in Figure \ref{hstfig}.

We find a remarkable, qualitative difference between the galaxies. Three are
resolved into hundreds of stars, as expected for a distance of $\sim 7$ Mpc. These
three galaxies (DF1, DF2, and DF3 in the \cite{Merritt2014} nomenclature) are
described in a companion paper (Danieli et al. 2016, submitted). However, the
remaining four (DF4, DF5, DF6 and DF7) are, at first inspection, \textit{undetected} in
our $HST$ imaging. The differences between the resolved and unresolved galaxies is illustrated in Figure \ref{ptsourcefig}. We select DF3 and DF4 as example galaxies, as they have nearly identical central surface brightness and size (Table \ref{paramstable}). Despite this, when we examine the images at full $HST$ resolution and the associated distributions of point sources detected with DOLPHOT (from Danieli et al.), DF3 emerges as a clear overdensity while DF4 appears to be an empty image. These two galaxies are representative of the differences between the two subsets of galaxies. Only after binning and smoothing can we infer the presence of large, diffuse objects in the images of DF4-DF7 $-$ although DF5-DF7 are better visually described as large scale background variations than obvious objects (see Figure \ref{hstfig}).

Figure \ref{cutoutsdf} displays greyscale images of all four unresolved galaxies (as observed with Dragonfly); color images of the central $100$ arcsec of each field, created from $HST$ $F606W$ and $F814W$ data, are shown in Figure \ref{cutoutshst}.

\subsection{CFHT}
\begin{figure*}[!t]
\begin{center}
\includegraphics[width=\textwidth]{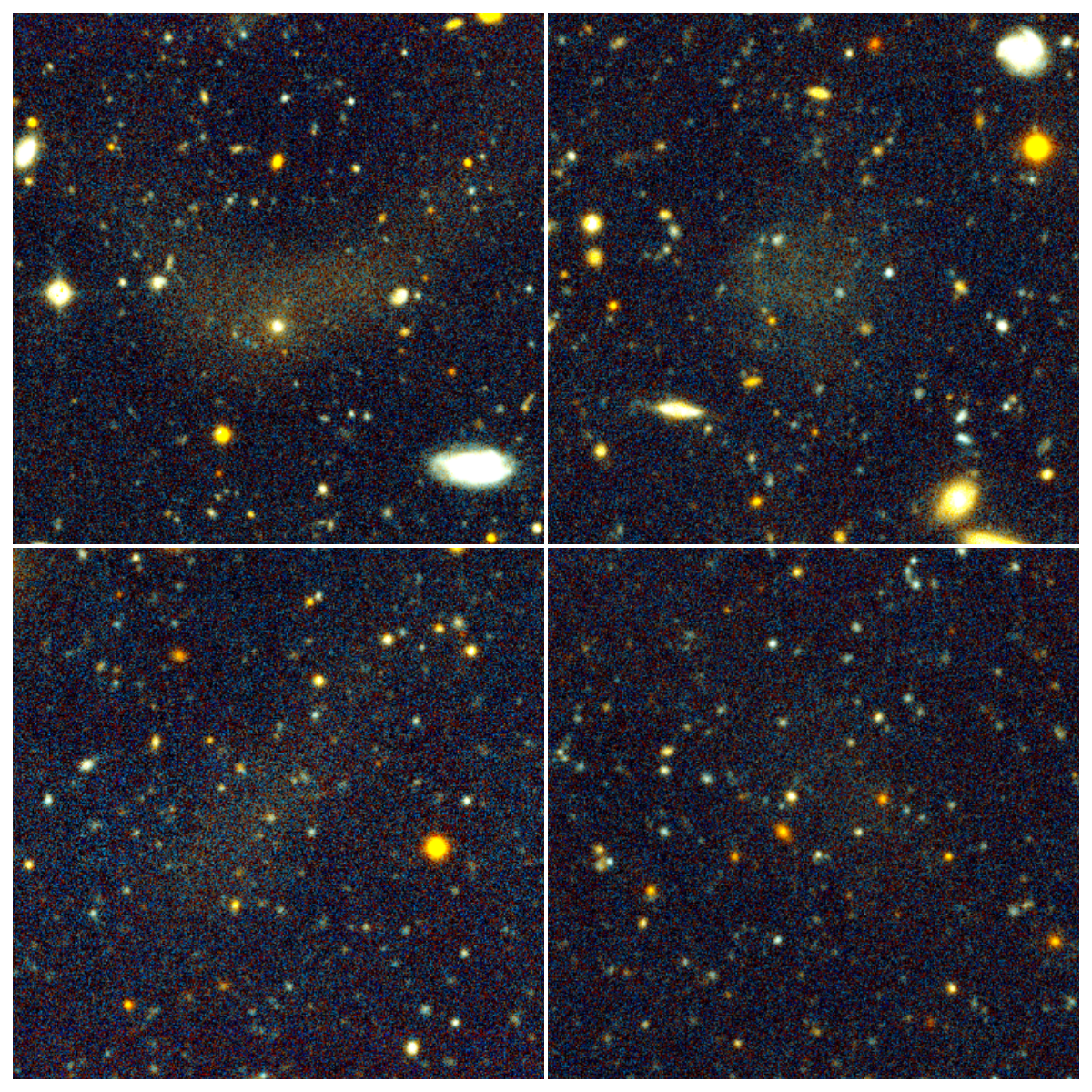}
\caption{CFHT pseudo-color images of the LSBGs (the area corresponds to the $100$
arcsec/side boxed regions in Figure \ref{cutoutsdf}), created from $g$ and $r$-band data. 
\label{cutoutscfht}}
\end{center}
\end{figure*}
The fact that these four galaxies do not resolve into stars raises the question of whether
our detection algorithm picked up noise peaks or artifacts in the Dragonfly data. This is
particularly a concern for DF6 and DF7, which are not significantly detected with $HST$
even after binning and smoothing. 

Fortunately, the M101 field has excellent additional datasets which settle this question.
The field has deep, public CFHT imaging obtained in the context of the 
Canada-France-Hawaii Telescope Lensing Survey \citep[CFHTLS][]{Heymans2012}. 
We obtained the reduced data from the CFHT archive;
the data were processed first with the standard \texttt{ELIXIR} \citep{Magnier2004}
software and then with the \texttt{THELI} data reduction pipeline \citep{Erben2005}. The
exposure times were 2500 s per filter.

The CFHT images of the four galaxies are shown in Figure \ref{cutoutscfht}.
All four objects are clearly detected, and a comparison between Figures \ref{cutoutsdf}
and \ref{cutoutscfht} shows that their morphologies are consistent between the two
datasets. 

\begin{figure*}[!t]
\begin{center}
\includegraphics[width=\textwidth]{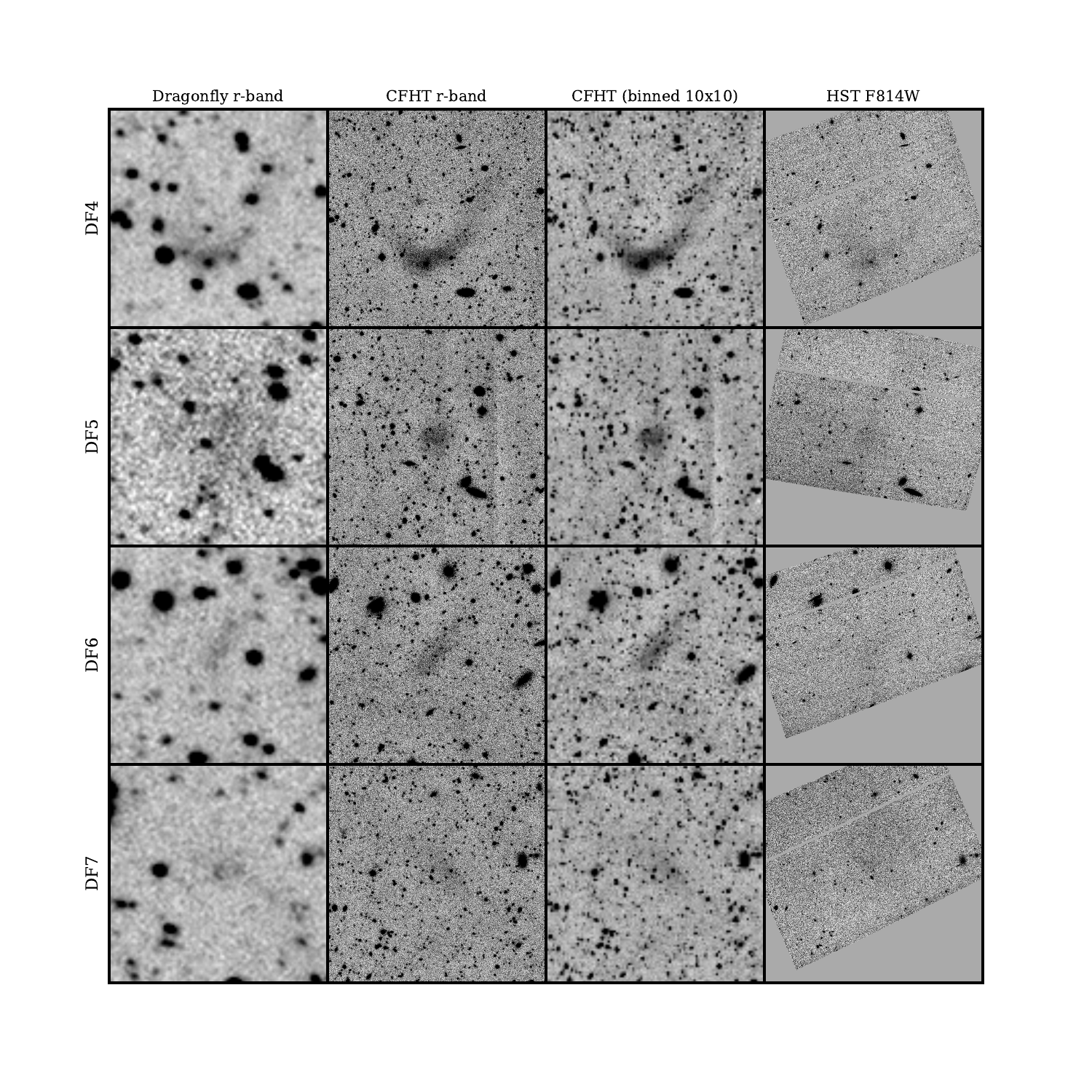}
\caption{Each row represents an individual LSBG (DF4-DF7). From
  left to right we show: the Dragonfly r-band cutout, as presented in
  \cite{Merritt2014}; the CFHT r-band cutout; the same CFHT r-band
  cutout binned to match the spatial resolution of Dragonfly; and the
  HST F814W cutout. Each box is 200 arcsec on a side. In the Dragonfly
  images, any overlapping stars were removed prior to analysis.
\label{datafig}}
\end{center}
\end{figure*}

Figure \ref{datafig} demonstrates that the apparent sizes of the galaxies
are somewhat dependent on the dataset, however. In particular, DF5 is the
largest of the Dragonfly-detected sample, but the smallest of the
CFHTLS-detected sample. The low spatial resolution of Dragonfly could,
in principle, result in an overestimated size if unresolved background
structures are erroneously included in the light of the galaxy in
question. Dragonfly is optimized to detect spatially-extended
emission, however, so another possibility is that the CFHTLS data is
less sensitive to the extremely diffuse outskirts of the galaxies.

\begin{figure}[!t]
\begin{center}
\includegraphics[scale=0.7]{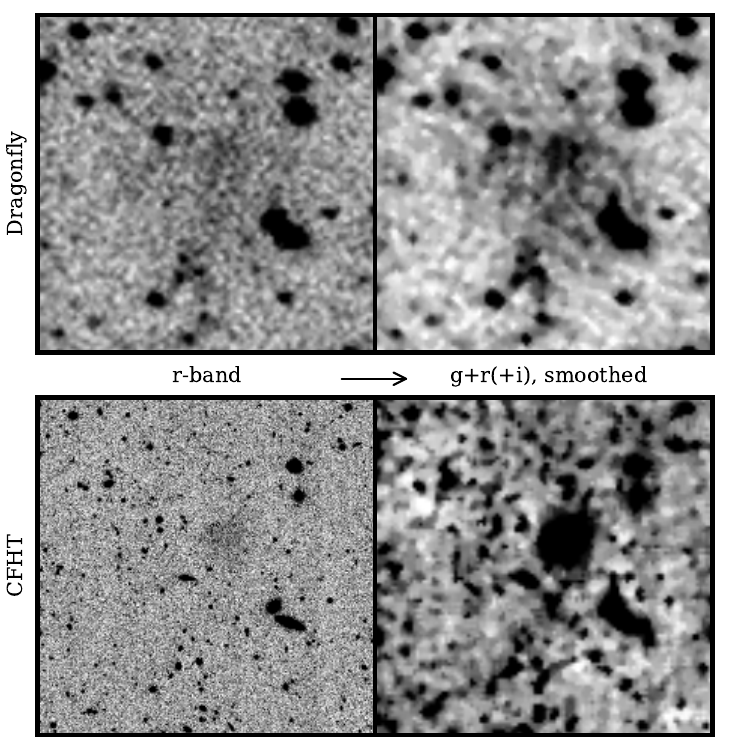}
\caption{The LSB galaxy DF5 is shown here in Dragonfly and CFHT
  data. The reduced data are shown in the left column. In the right
  column, the Dragonfly image has been median-smoothed (3x3 pixels),
  and the CFHT image has been first binned to match the spatial
  resolution of Dragonfly and then median-smoothed in the same
  way. While the size of DF5 is apparently smaller in the CFHT data
  than in Dragonfly in both cases, in the binned/smoothed image faint
  emission can be seen extending out much farther than the reduced
  image reveals. Additionally, CCD defects in the CFHT data can be
  seen in both images. We therefore use only the Dragonfly data from
  this point onward.
\label{binnedfig}}
\end{center}
\end{figure}

As a test, we bin the CFHT data to match the spatial resolution of
Dragonfly (approximately $10\times 10$ pixels) and then median-smooth with a
kernel of $3$ pixels to enhance the low surface brightness outskirts
(Figure \ref{binnedfig}). The Dragonfly data were also median-smoothed
for consistency. After binning and smoothing, DF5 is still smaller in
the CFHT image than in the Dragonfly image. However, we can see that
this process has indeed increased the flux at large radii, indicating that the most
likely culprit is either the surface brightness limits of the CFHT data or
possibly an overestimate of the background level during sky subtraction. We
therefore adopt the values of surface brightness, structure and
sizes measured in the Dragonfly data for the remainder of this paper.

\begin{deluxetable*}{ccccccccccccc}
\tabletypesize{\footnotesize}
\tablecolumns{13}
\tablewidth{0pt}
\tablecaption{Physical Properties of the UDGs \label{paramstable}}
\tablehead{
\colhead{ID} & \colhead{$\alpha$} & \colhead{$\delta$} & \colhead{$g$\,\tablenotemark{a}} &
\colhead{$\mu_{0,g}$\,\tablenotemark{b}} & \colhead{$\mu_{e,g}$\,\tablenotemark{c}} &
\colhead{$g-r$} & \colhead{$r_{e}$\, \tablenotemark{d}} & 
\colhead{$r_{e}$\,\tablenotemark{e}} &
\colhead{$n$\,\tablenotemark{f}} & \colhead{$b/a$\,\tablenotemark{g}} \\
\colhead{} & \colhead{(J2000)} & \colhead{(J2000)} & \colhead{} &
\colhead{} & \colhead{} & \colhead{} & \colhead{} & \colhead{} & \colhead{} & \colhead{}
}
\startdata
M101-DF4 & 211.88932 & 54.710178 & 18.8 $\pm$ 0.3 & 26.8 $\pm$ 0.4 & 27.9 $\pm$ 0.2 & 0.6 $\pm$ 0.4 & $28 \pm 7$ & $3.6 \pm 0.9$ & 0.7 $\pm$ 0.3 & 0.6 $\pm$ 0.1 \\ 
M101-DF5 & 211.11709 & 55.616788 & 18.0 $\pm$ 0.2 & 27.4 $\pm$ 0.3 & 28.0 $\pm$ 0.2 & 0.4 $\pm$ 0.4 & $38 \pm 7$ & $4.9 \pm 0.9$ & 0.4 $\pm$ 0.2 & 0.8 $\pm$ 0.1 \\ 
M101-DF6 & 212.07927 & 55.190214 & 20.1 $\pm$ 0.4 & 27.5 $\pm$ 1.1 & 27.8 $\pm$ 0.4 & 0.4 $\pm$ 0.5 & $22 \pm 8$ & $2.9 \pm 1.0$ & 0.3 $\pm$ 0.8 & 0.3 $\pm$ 0.1 \\ 
M101-DF7 & 211.45134 & 55.132899 & 20.4 $\pm$ 0.6 & 27.7 $\pm$ 1.6 & 28.7 $\pm$ 0.6 & 0.9 $\pm$ 0.8 & $20 \pm 9$ & $2.6 \pm 1.1$ & 0.6 $\pm$ 1.0 & 0.5 $\pm$ 0.2 \\ 
\enddata
\tablenotetext{a}{Integrated apparent magnitude, calibrated to SDSS.}
\tablenotetext{b}{Central surface brightness, in mag arcsec$^{-2}$.}
\tablenotetext{c}{Effective surface brightness, in mag arcsec$^{-2}$.}
\tablenotetext{d}{Effective radius, in arcsec.}
\tablenotetext{e}{Effective radius, in kpc, assuming a distance of 27 Mpc.}
\tablenotetext{f}{Sersic index.}
\tablenotetext{g}{Axis ratio.}
\tablecomments{Structural parameters were computed using GALFIT, from
  a stack of Dragonfly \(g\)- and \(r\)-band images.}
\end{deluxetable*}

\section{Distance Measurements}

\subsection{A lower distance limit based on $HST$ images}
For nearby galaxies, the Tip of the Red Giant Branch (TRGB) $I$-band luminosity
provides a reliable distance estimator as it corresponds to a constant value of
$M_{I}^{TRGB} \sim -4$ mag \citep[e.g.,][]{Gallart2005}. This technique relies on the ability
to detect individual RGB stars, however, and we can therefore use the fact that we are
\textit{unable} to resolve these galaxies with $HST$ to place lower limits on their distances.

In Danieli et al. (2016, submitted) we use the publicly available DOLPHOT software
package \citep{Dolphin2000} in combination with the TRGBTOOL software \citep{Makarov2006}
to determine the limiting magnitude for each resolved LSBG
(DF1, DF2, and DF3). In brief, we create an artificial star list based on the photometry
of each LSBG, place the stars into the $HST$ images, and use the resulting DOLPHOT outputs
to measure the completeness as a function of magnitude.

We find that the data reach $50\%$ completeness at $27.2^{+0.04}_{-0.03}$ magnitude in
$I$-band ($F814W$). This value is the average of the results for DF1, DF2, and DF3; the
error bars encompass the full variation between the three $HST$ fields.
To first order, then, we can infer a minimum distance modulus of $31.2$
mag from an undetected TRGB, corresponding to a lower distance limit of $17.5$ Mpc. We note that for a galaxy at the distance of M101, we would expect to observe the TRGB at an $I$-band magnitude of $\sim 25$.

\begin{figure*}[!t]
\begin{center}
\includegraphics[width=\textwidth]{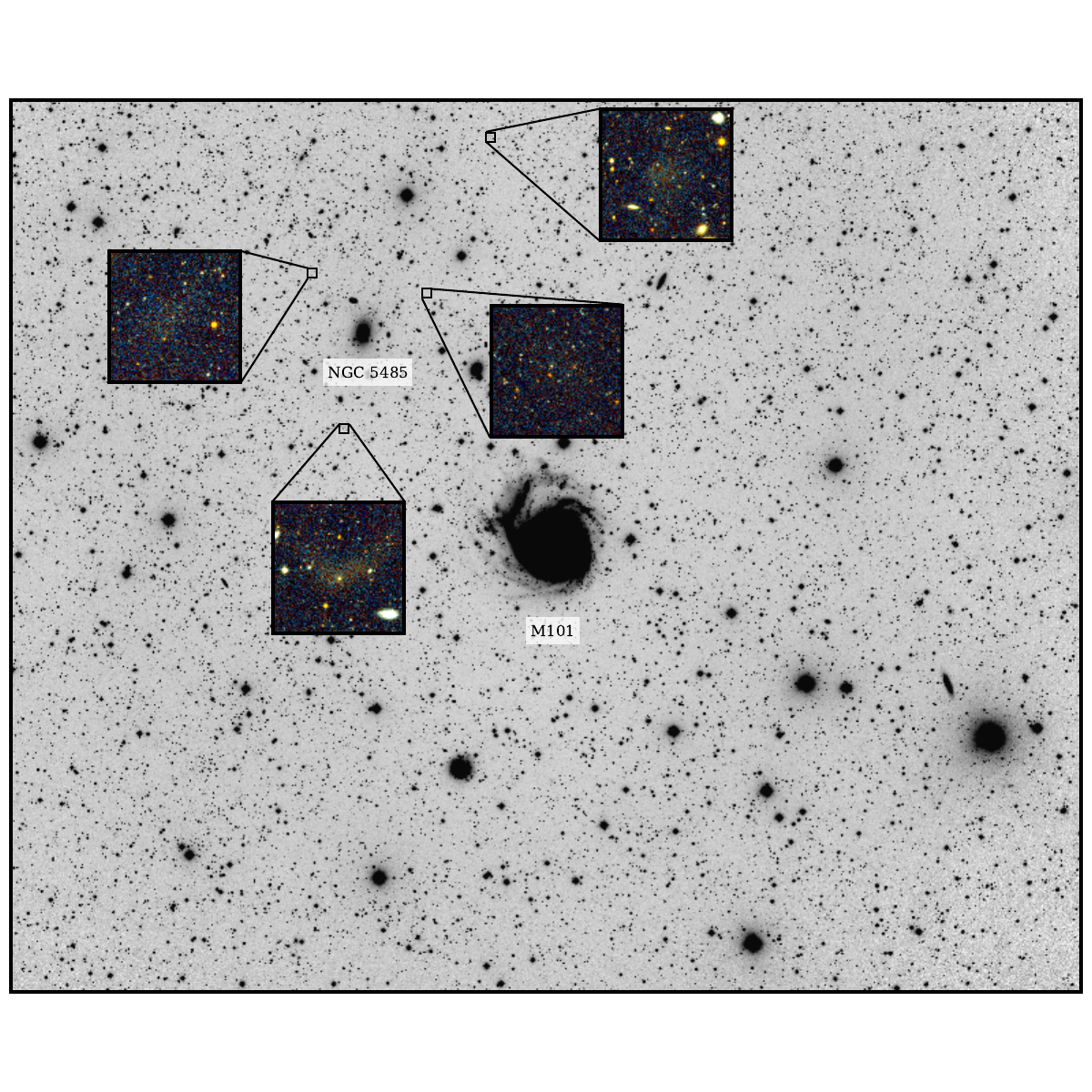}
\caption{The Dragonfly field of view is shown in greyscale, with zoom insets showing the
locations of the four LSBGs. The color images were created using CFHT g- and r-band data.
\label{fovcolor}}
\end{center}
\end{figure*}

\subsection{Association with the NGC 5485 group}
From the lower bounds on the distances to the galaxies, we infer that
they cannot be members of the M101 group, and we
turn to the surrounding field for clues of their local environment. 

Figure \ref{fovcolor} shows the full Dragonfly field centered on M101 where these
galaxies were discovered. The zoomed panels show CFHT color images of the galaxies
and highlight their positions in the field. As previously noted by \cite{Merritt2014},
all four lie to the East of M101; they appear to be distributed nearly uniformly around
NGC 5485, a massive elliptical galaxy at a distance of $\sim 27$ Mpc \citep{Tully2016}.

The galaxies appear to be consistent with the projected locations of NGC 5485 and
its associated
group, which has NGC 5473 as its brightest member \cite{Tully2015}. However, the 
\cite{Tully2015} group catalog is optimized for $3,000 \lesssim v_{r} \lesssim 10,000$ km
s$^{-1}$ and suffers from uncertainties due to
large peculiar velocities for $v_{r} \lesssim 3,000$ km s$^{-1}$ (the
regime relevant for all nearby groups of interest) as well as incompleteness. We therefore
supplement it with the \cite{MK2011} groups catalog, which is more
accurate for nearby groups \citep{MK2011,Tully2015}. Figure \ref{grouppos} displays a 
zoom-in on the members of the NGC 5473 group, with symbol sizes of group members scaled by
their absolute $B$ magnitude. The positions of the galaxies, denoted by star symbols, are 
centered on and distributed throughout the group.

\section{Results}

\subsection{Sizes}
Given their projected proximity to the NGC 5485 group, we adopt a distance
of $27$ Mpc for the four unresolved LSBGs. At this distance, their physical sizes
range from $2.6 \pm 1.1$ to $4.9 \pm 0.9$ kpc. 

These sizes, in combination with the low central surface brightnesses, are
large enough to allow us to classify the unresolved LSBGs as UDGs, a population of galaxies
empirically defined by \cite{vanDokkum2015a} to have extremely low central surface
brightness ($\mu_{g,0} > 24$ mag arcsec$^{-2}$) and large sizes ($R_{e}
> 1.5$ kpc). 

\begin{figure*}[!t]
\begin{center}
\includegraphics[width=\textwidth]{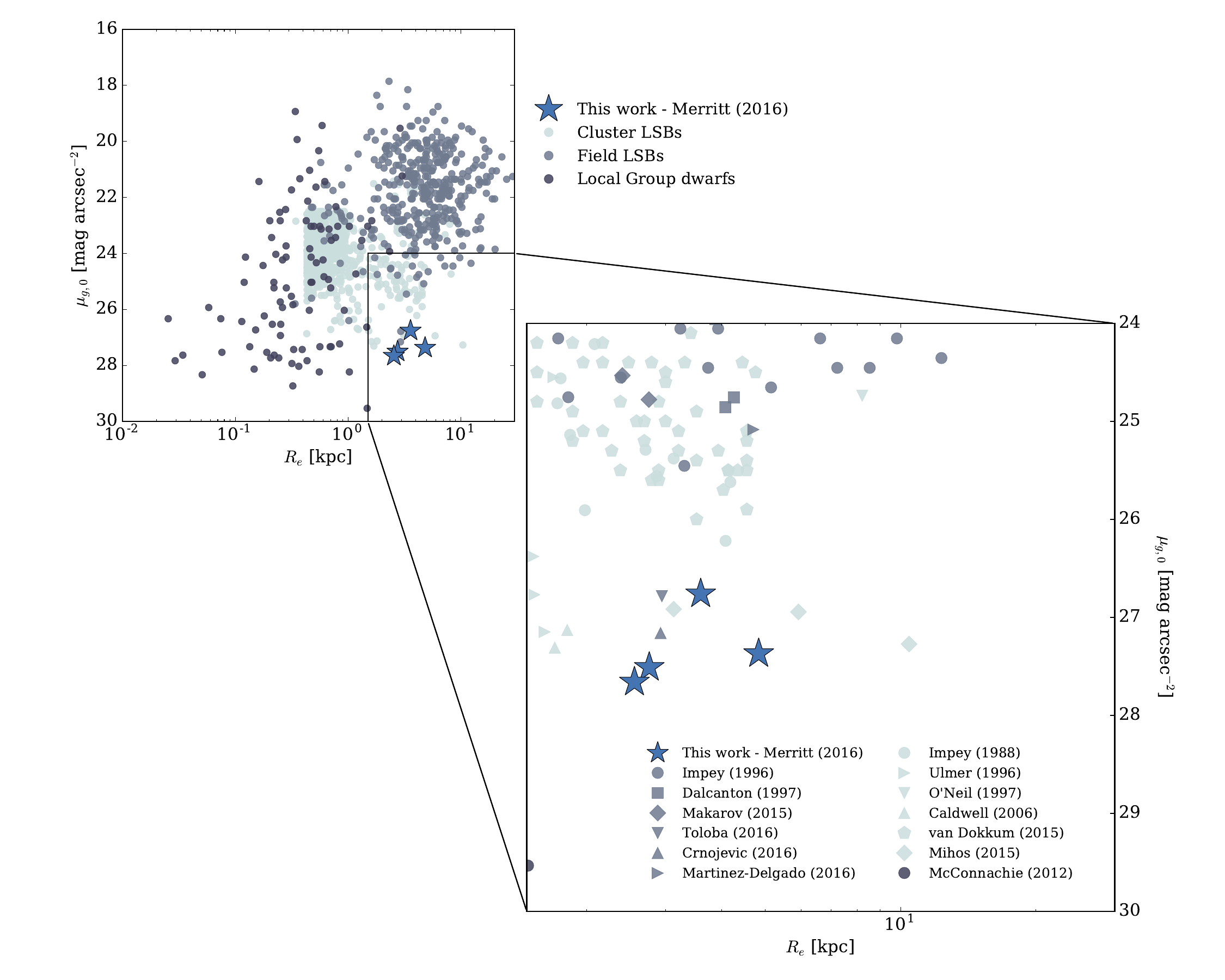}
\caption{The four group UDGs in the context of other known LSBGs.
  Upper left: the surface brightness - effective radius
  plane for LSBs identified in clusters, in the field, and in the
  Local Group. Lower right: a zoom-in on the region empirically
  defined to host UDGs. The majority of objects here are found in
  clusters, although a few are found in lower density environments
  and one (And XIX) exists in the Local Group. Where necessary, we
  have converted reported paramters to a cosmology with
  $H_{0}=70$ km s$^{-1}$ Mpc$^{-1}$, and converted from scale length
  to effective radii. Any required color transformations were done
  using equations from \cite{Blanton2007} and \cite{Fukugita1996}. If
  a study listed an assumed (rather than measured) distance,
  we assumed the same. We note that a handful of studies shown in the
  upper left plot do not contain any candidate UDGs $-$ these include
  \cite{Bothun1991}, \cite{McGaugh1994}, \cite{deBlok1995} and
  \cite{Davies2015}.
\label{theplot}}
\end{center}
\end{figure*}

We provide a graphical summary of known (available) low surface brightness galaxy
catalogs in Figure \ref{theplot}. We note that classically,
LSBGs are defined on the basis of the central surface brightness of the disk,
irrespective of the presence of a bulge component. In order to provide the most fair
comparison with our sample, we consider only bulgeless LSBGs with $n \leq 1$.
The zoom box highlights the empirically defined UDG
region ($\mu_{0,g} > 24$ mag arcsec$^{-2}$ and $R_{e} > 1.5$ kpc). The galaxies
presented here (shown in Figure \ref{theplot} as large green symbols) are a
sample of UDGs observed in a group environment, and are among the most
extreme objects of their class in terms of surface brightness.

\subsection{Colors and morphologies \label{morph}}
The morphologies of these four UDGs are highly nonuniform, with 
axis ratios ($0.3 \leq b/a \leq 0.9$). The colors and structural parameters 
as measured from Dragonfly photometry were reported in \cite{Merritt2014},
and we summarize these along with the physical sizes of each galaxy in Table
\ref{paramstable}. The average $g-r$ color of the sample is $\langle g-r \rangle = 0.57$, similar to the average color of the low luminosity end of the sample studied by
\cite{vanderBurg2016}.


The group UDGs display a noteworthy
degree of morphological diversity when compared to the cluster UDGs. The latter are
predominantly round \citep[e.g.][]{vanDokkum2015a,Mihos2015}, whereas
the galaxies presented here have a large range of ellipticities (see
Table \ref{paramstable}). One galaxy (DF4) has a boomerang-like shape,
and is possibly undergoing tidal stripping. The largest, DF5, is the roundest
of the sample; the remaining two (DF6 and DF7) are elongated.
Despite these differences, however, the surface brightness
profiles of group and cluster UDGs alike fall off exponentially,
occasionally featuring a central depression \citep[that is, $n < 1$,
as measured by][see also Table \ref{paramstable}]{Merritt2014,vanDokkum2015a}. 

There are no apparent trends between color and morphology in this small sample.

\begin{figure}[!t]
\begin{center}
\includegraphics[scale=0.5]{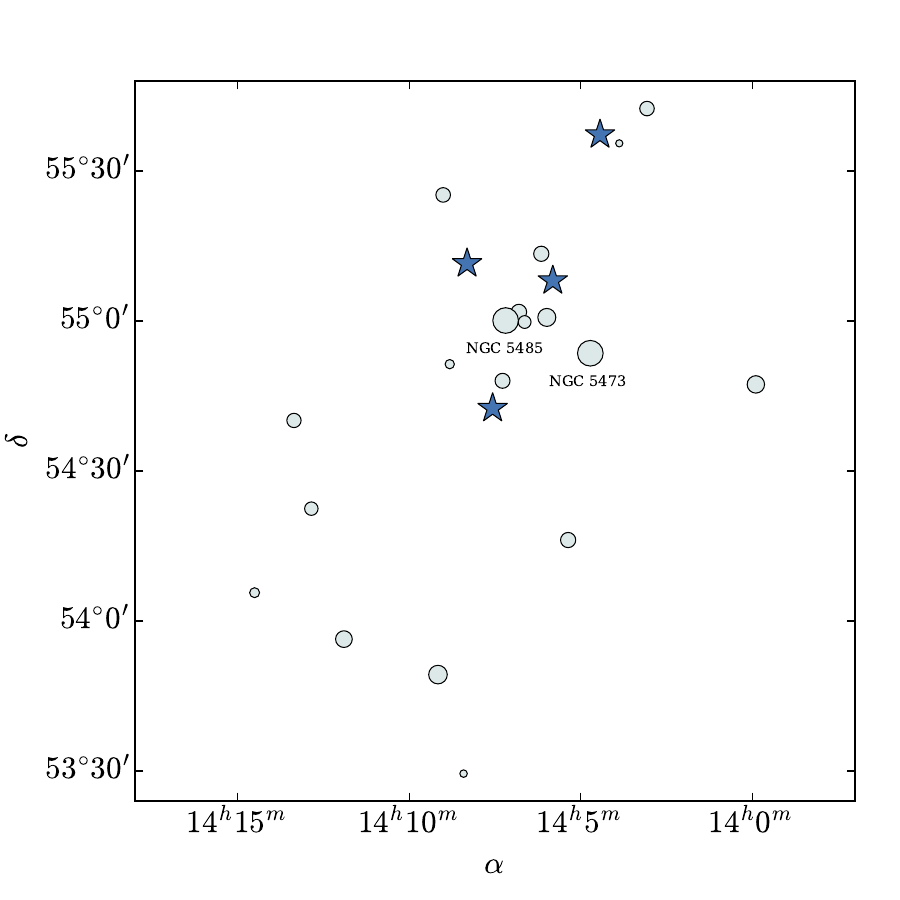}
\caption{The members of the NGC 5485 group \citep{MK2011}, with symbol sizes
  scaled by absolute B magnitude \citep{LEDA}. The locations of the four LSBGs
  are shown as well (blue stars; no luminosity scaling); the projected positions are
  consistent with group membership.
\label{grouppos}}
\end{center}
\end{figure}

\begin{figure*}[!t]
\begin{center}
\includegraphics[width=\textwidth]{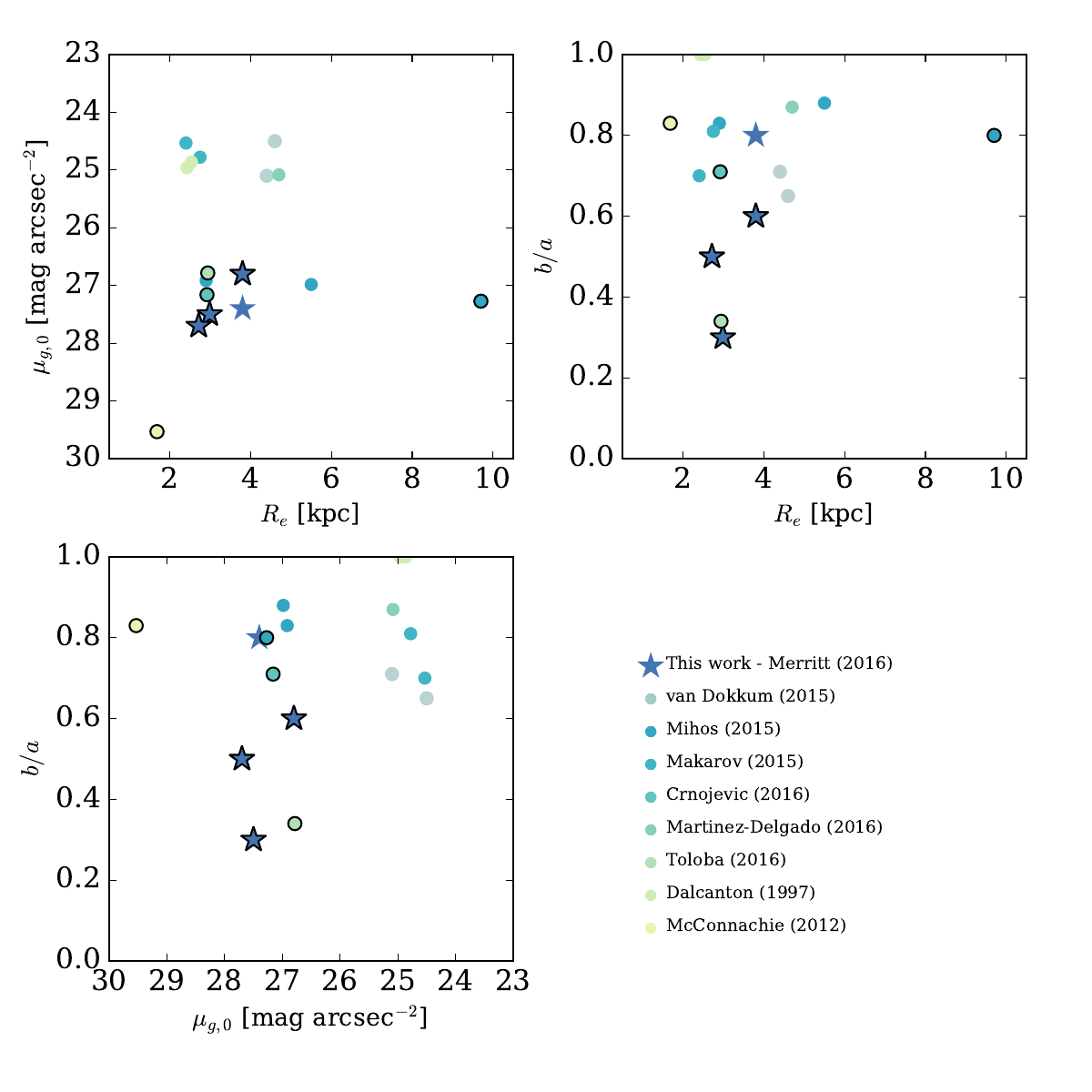}
\caption{A comparison between the central surface brightnesses, sizes, and axis ratios
of known UDGs \textit{with measured distances} (or upper limits on distance). Literature
points with black circles indicate cases where the authors describe the galaxy as plausibly
disrupting. In general, quoted surface brightnesses in the literature were in $V$-band;
we used reported $B-V$ colors to convert to $g$-band where possible, and the average of
reported colors ($\langle B-V \rangle = 0.72$) otherwise (necessary for data points
from \cite{Toloba2016}, \cite{MartinezDelgado2016}, and VLSB-C from \cite{Mihos2015}).
An exceptions to this, however, is the sample \cite{McConnachie2012}, for which we use
$\langle B-V \rangle = 0.63$
\citep[the mean color of the brighter Local Group dwarfs, as measured by][]{Mateo1998}.
The axis ratio for VLSB-A \citep[from][]{Mihos2015} is an upper limit.
\label{udgprops}}
\end{center}
\end{figure*}

\section{Discussion}

\subsection{UDGs outside of the cluster environment}
The minimum distance to the four unresolved LSBGs places strong lower
limits on their sizes and confirms their status as UDGs; the projected
spatial distribution of the galaxies strongly suggests that they are
members of the NGC 5473 group at a distance of $\sim 27$ Mpc, providing further evidence
that UDGs are not a phenomenon that is exclusive to cluster
environments. 

Thus far, the growing observational census for the UDG population
suggests that UDGs are preferentially 
associated with massive early-type galaxies, with the presence of cluster UDGs in
particular becoming increasingly well documented \citep{Impey1988,ONeil1997a,Bothun1991,Ulmer1996,Caldwell2006,vanDokkum2015a,Davies2015,Mihos2015,Koda2015,Munoz2015,vanderBurg2016,Yagi2016}.
Additional evidence in support of this picture comes from
\cite{Munoz2015}, who found that the non-nucleated LSBG population in
Fornax has a projected spatial distribution that clusters around the
locations of giant ellipticals (we note however, that their sample
contains very few UDGs); from \cite{vanderBurg2016}, who showed that 
the number density of UDGs in clusters is correlated with cluster
halo mass; and from \cite{RomanTrujillo2016} who reported
that UDGs tend to be located within large-scale structures.

Examples of UDGs with either established \citep{Makarov2015,MartinezDelgado2016}
or implied \citep{Dalcanton1997,ONeil1997a} lower density environments do exist, however.
And, recently, they have even been associated with spiral galaxies $-$ specifically
in the cases of And XIX in the Local Group \citep{McConnachie2012} and Scl-MM-Dw2 in the
NGC 253 group \citep{Toloba2016}.

To quantify the population of UDGs in group environments, we compare the number of known UDGs to the number of bright ($M_{B} < -17$) group members. We obtain group membership information from \cite{MK2011}, and absolute $B$-band magnitudes from \cite{LEDA}.
The $4$ UDGs presented here, relative to the $6$ bright members of the NGC 5473 group,
constitute a relatively rich UDG population when compared to the Local Group
($1$ known UDG per $6$ bright members). The NGC 253 and Cen A groups
each have $1$ known UDG as well, compared to $2$ and $4$
bright members, respectively. Although it is unclear based on this small sample whether we should necessarily expect a tight correlation
between the number of bright members and the number of UDGs in a given environment, these
numbers do suggest that there should be at least one observable UDG per nearby group.

\subsection{Structural stability}
Key questions regarding the UDG population include: Is this a stable
population, or are we witnessing their disruption and the subsequent
buildup of intragroup and intracluster light? If they are stable, how
can such diffuse galaxies survive?

The three cluster UDGs $-$ DF44, DF17 and VCC 1287 $-$ that have been studied in
detail thus far have all been shown to have very high mass-to-light ratios in their
central regions \citep{vanDokkum2016,Beasley2016,BeasleyTrujillo2016}. All three have
globular cluster systems (and, in the case of DF44, a stellar velocity dispersion) that are
typical of galaxies in relatively massive dark matter halos, and a stellar component that is
significantly underluminous. In combination with their (undisturbed) spheroidal morphologies
and old stellar populations, these galaxies can potentially be thought of as
``failed'' $L_{*}$ or dwarf galaxies that lost their gas early on
\citep[e.g., by ram pressure stripping,][]{GunnGott1972}. 

An alternate explanation is that some UDGs are either the descendants of classical (bulgeless)
LSBGs, which are thought to reside in massive, low density dark matter halos
\citep[e.g.][]{Dalcanton1995,deBlok1996} ; or, possibly, ``almost-darks''
\citep[e.g.][]{Cannon2015} that fell into denser environments. 
In this scenario, the colors and morphologies of the otherwise slowly-evolving LSBGs
\citep{vandenHoek2000} could be explained by an accelerated evolution induced by
interactions with neighbors in high density environments \citep{ONeil1997b}. 
\cite{Gnedin2003} showed that although the most diffuse and extended LSBGs are fated to
disrupt in cluster environments \citep[see also][]{Moore1999}, others will simply lose the
majority of their halo mass (and up to $\sim 20$\% of their stellar mass) and transform
into spheroidal systems with low central surface densities and large effective radii. 
This particular scenario is ruled out for DF44, DF17 and VCC 1287, as LSBGs have been
shown to contain ``normal'' globular cluster populations when compared to HSBGs at similar
luminosity \citep{Villegas2008} and if the galaxy had lost (and redistributed) enough mass
to transform into a spheroidal UDG, it should have also lost a substantial fraction of its
globular clusters. It could, however, potentially explain UDGs with depleted
globular cluster systems.

It is worth noting that the Local Group analog of the group and cluster UDGs is And XIX
\citep{McConnachie2008,McConnachie2012}, a dwarf satellite of M31 that is known to be
disrupting \citep{Collins2013}. The UDGs recently identified around Cen A and NGC 253 
by \cite{Crnojevic2016} and \cite{Toloba2016} show evidence for disruption as well, in the
form of extended, prominent tidal tails and an elongation towards the the central massive
galaxy, respectively. \cite{Mihos2015} also reported that one of the three UDGs that they
identified in the Virgo cluster (VLSB-A) is disrupting, as indicated by its elongated
structure. Additionally, the apparent absence of UDGs in the central regions of the Coma
cluster \cite{vanDokkum2015a} could be straightforwardly explained if UDGs were unable to
survive at small cluster-centric distances; and \cite{vanderBurg2016} find that the
radial number density distribution of UDGs in eight clusters is consistent with a
lack of UDGs within $300$ kpc of the cluster centers. 
It is also worth pointing out that NGC 5485 itself is enveloped by several tidal
features \cite[e.g.,][]{Karachentsev2015}.

If the four group UDGs presented here are unstable and tidally
disrupting, then we should expect to see this reflected in their morphologies. As discussed
in Section \ref{morph}, DF4 is highly distorted, with a boomerang-like appearance. DF6 and
DF7 are elongated, although not obviously pointed towards the group center (this is not ruled
out, however, considering projection effects). This would appear to be at odds with the
round morphologies and apparently regular appearance of cluster UDGs
\citep[the average axis ratio for the Coma sample is $0.7$;][]{vanDokkum2015a}. 
A detailed analysis morphologies of cluster UDGs is beyond the scope of this work, but we are investigating this in a separate paper (Mowla et al., in preparation). 

In Figure \ref{udgprops}, we show the size, axis ratio, and central surface brightness of
every known UDG reported here as well as in the literature.
Unlike Figure \ref{theplot}, this time we show
\textit{only} those UDGs with measured distances (or upper limits); that is, if the
discovery paper assumed a distance or association with a background object, we do not include
those data points. The UDGs in this work and in the literature are represented by stars and
circles, respectively, and symbols with black borders highlight the UDGs that have been
reported to be disrupting. We note that there is an apparent clustering of the disrupting UDGs
seen in the lower left panel (in the axis ratio $-$ surface brightness plane), such that they
tend to inhabit the lowest surface brightness regime ($\mu_{g,0} > 26.5$ mag arcsec$^{-2}$)
and, while they cover a wide range in axis ratios, the majority have $b/a \lesssim 0.7$.

Intruigingly, the roundest UDG in our small sample $-$ DF5 $-$ also has the largest
projected distance from the group center (see Figures \ref{fovcolor} and \ref{grouppos}). If DF5 is \textit{not} associated with the NGC 5485 group, it could be an example of an isolated UDG. The identification of a population of field UDGs would provide critical clues to their formation. 

Regardless of whether UDGs are stable or on the brink of disruption, the fact that we
observe these galaxies residing in groups suggests that group pre-processing may play a
significant role in their formation and subsequent arrival in clusters
\citep{Wetzel2015,Vijayaraghavan2013}; in this case we expect that future
searches of nearby groups will turn up comparable populations of
UDGs.

\section{Conclusions}
We have presented follow-up $HST$/ACS observations of four LSBGs discovered in
\cite{Merritt2014}; the other three resolved galaxies are presented in
Danieli et al. (2016, submitted). We placed lower distance limits of $17.5$ Mpc 
based on the lack of resolved stars in the $HST$ imaging. Given that the distance to
the UDGs rules out the possibility of a M101 group membership, we consider that they are
instead part of the background NGC 5473 / 5485 group, located at $\sim 27$ Mpc. 
The projected positions of the UDGs are distributed evenly around the center
of the group, consistent with this picture. At this
distance, the galaxies have effective radii of $2.6-4.9$ kpc, and the large physical sizes
combined with the low central surface brightnesses
($\mu_{g,0} = 25.6 - 27.7$ mag arcsec$^{-2}$) qualifies them as UDGs.

Moving forward, it will be critical to expand the census of known UDGs even further,
particular to lower density environments. The identification of group UDGs is perhaps not
surprising given their apparent ubiquity in clusters, although finding four in a single group is remarkable given that there are only three (with measured distances) known in the
Virgo cluster thus far. The morphologies of the group UDGs are, on average, more complex
than the morphologies of cluster UDGs. If a significant fraction of the UDG population
is comprised of puffed-up dwarfs, then it is possible that we are witnessing that
transformation happening.

\acknowledgments
We thank the anonymous referee for helpful comments that improved the paper.
Support from NSERC, NSF grant AST-1312376, grant HST-GO-13682,
and from the Dunlap Institute (funded by the David Dunlap Family) is gratefully acknowledged.
All authors thank the staff at New Mexico Skies Observatory for their support and assistance;
IDK and LNM acknowledge the support of the Russian Science Foundation grant 14-12-00965. AM
thanks the BS group for useful discussions.


\begin{thebibliography}{66}
\expandafter\ifx\csname natexlab\endcsname\relax\def\natexlab#1{#1}\fi

\bibitem[{{Abraham} \& {van Dokkum}(2014)}]{Abraham2014}
{Abraham}, R.~G., \& {van Dokkum}, P.~G. 2014, \pasp, 126, 55

\bibitem[{{Adami} {et~al.}(2006){Adami}, {Scheidegger}, {Ulmer}, {Durret},
  {Mazure}, {West}, {Conselice}, {Gregg}, {Kasun}, {Pell{\'o}}, \&
  {Picat}}]{Adami2006}
{Adami}, C., {Scheidegger}, R., {Ulmer}, M., {et~al.} 2006, \aap, 459, 679

\bibitem[{{Amorisco} \& {Loeb}(2016)}]{AmoriscoLoeb2016}
{Amorisco}, N.~C., \& {Loeb}, A. 2016, \mnras, 459, L51

\bibitem[{{Beasley} {et~al.}(2016){Beasley}, {Romanowsky}, {Pota}, {Navarro},
  {Martinez Delgado}, {Neyer}, \& {Deich}}]{Beasley2016}
{Beasley}, M.~A., {Romanowsky}, A.~J., {Pota}, V., {et~al.} 2016, \apjl, 819,
  L20

\bibitem[{{Beasley} \& {Trujillo}(2016)}]{BeasleyTrujillo2016}
{Beasley}, M.~A., \& {Trujillo}, I. 2016, ArXiv:1604.08024

\bibitem[{{Blanton} \& {Roweis}(2007)}]{Blanton2007}
{Blanton}, M.~R., \& {Roweis}, S. 2007, \aj, 133, 734

\bibitem[{{Bothun} {et~al.}(1991){Bothun}, {Impey}, \& {Malin}}]{Bothun1991}
{Bothun}, G.~D., {Impey}, C.~D., \& {Malin}, D.~F. 1991, \apj, 376, 404

\bibitem[{{Bothun} {et~al.}(1987){Bothun}, {Impey}, {Malin}, \&
  {Mould}}]{Bothun1987}
{Bothun}, G.~D., {Impey}, C.~D., {Malin}, D.~F., \& {Mould}, J.~R. 1987, \aj,
  94, 23

\bibitem[{{Caldwell}(2006)}]{Caldwell2006}
{Caldwell}, N. 2006, \apj, 651, 822

\bibitem[{{Cannon} {et~al.}(2015){Cannon}, {Martinkus}, {Leisman}, {Haynes},
  {Adams}, {Giovanelli}, {Hallenbeck}, {Janowiecki}, {Jones}, {J{\'o}zsa},
  {Koopmann}, {Nichols}, {Papastergis}, {Rhode}, {Salzer}, \&
  {Troischt}}]{Cannon2015}
{Cannon}, J.~M., {Martinkus}, C.~P., {Leisman}, L., {et~al.} 2015, \aj, 149, 72

\bibitem[{{Collins} {et~al.}(2013){Collins}, {Chapman}, {Rich}, {Ibata},
  {Martin}, {Irwin}, {Bate}, {Lewis}, {Pe{\~n}arrubia}, {Arimoto}, {Casey},
  {Ferguson}, {Koch}, {McConnachie}, \& {Tanvir}}]{Collins2013}
{Collins}, M.~L.~M., {Chapman}, S.~C., {Rich}, R.~M., {et~al.} 2013, \apj, 768,
  172

\bibitem[{{Crnojevi{\'c}} {et~al.}(2016){Crnojevi{\'c}}, {Sand}, {Spekkens},
  {Caldwell}, {Guhathakurta}, {McLeod}, {Seth}, {Simon}, {Strader}, \&
  {Toloba}}]{Crnojevic2016}
{Crnojevi{\'c}}, D., {Sand}, D.~J., {Spekkens}, K., {et~al.} 2016, \apj, 823,
  19

\bibitem[{{Dalcanton} {et~al.}(1997){Dalcanton}, {Spergel}, {Gunn}, {Schmidt},
  \& {Schneider}}]{Dalcanton1997}
{Dalcanton}, J.~J., {Spergel}, D.~N., {Gunn}, J.~E., {Schmidt}, M., \&
  {Schneider}, D.~P. 1997, \aj, 114, 635

\bibitem[{{Dalcanton} {et~al.}(1995){Dalcanton}, {Spergel}, \&
  {Summers}}]{Dalcanton1995}
{Dalcanton}, J.~J., {Spergel}, D.~N., \& {Summers}, F. 1995, ArXiv Astrophysics
  e-prints

\bibitem[{{Danieli} {et~al.}(2016, submitted){Danieli}, {van Dokkum}, {Merritt}, {Abraham}, {Zhang}, {Karachentsev} \& {Makarova}}]{Danieli2016}
{Danieli}, S., {van Dokkum}, P.~G., {Merritt}, A., {Abraham}, R.~G., {Zhang}, J., {Karachentsev}, I.~D., \& {Makarova}, L.~N. 2016, submitted

\bibitem[{{Davies} {et~al.}(2016){Davies}, {Davies}, \& {Keenan}}]{Davies2015}
{Davies}, J.~I., {Davies}, L.~J.~M., \& {Keenan}, O.~C. 2016, \mnras, 456, 1607

\bibitem[{{de Blok} \& {McGaugh}(1996)}]{deBlok1996}
{de Blok}, W.~J.~G., \& {McGaugh}, S.~S. 1996, \apjl, 469, L89

\bibitem[{{de Blok} {et~al.}(1995){de Blok}, {van der Hulst}, \&
  {Bothun}}]{deBlok1995}
{de Blok}, W.~J.~G., {van der Hulst}, J.~M., \& {Bothun}, G.~D. 1995, \mnras,
  274, 235

\bibitem[{{Disney}(1976)}]{Disney1976}
{Disney}, M.~J. 1976, \nat, 263, 573

\bibitem[{{Dolphin}(2000)}]{Dolphin2000}
{Dolphin}, A.~E. 2000, \pasp, 112, 1383

\bibitem[{{Erben} {et~al.}(2005){Erben}, {Schirmer}, {Dietrich}, {Cordes},
  {Haberzettl}, {Hetterscheidt}, {Hildebrandt}, {Schmithuesen}, {Schneider},
  {Simon}, {Deul}, {Hook}, {Kaiser}, {Radovich}, {Benoist}, {Nonino}, {Olsen},
  {Prandoni}, {Wichmann}, {Zaggia}, {Bomans}, {Dettmar}, \&
  {Miralles}}]{Erben2005}
{Erben}, T., {Schirmer}, M., {Dietrich}, J.~P., {et~al.} 2005, Astronomische
  Nachrichten, 326, 432

\bibitem[{{Fukugita} {et~al.}(1996){Fukugita}, {Ichikawa}, {Gunn}, {Doi},
  {Shimasaku}, \& {Schneider}}]{Fukugita1996}
{Fukugita}, M., {Ichikawa}, T., {Gunn}, J.~E., {et~al.} 1996, \aj, 111, 1748

\bibitem[{{Gallart} {et~al.}(2005){Gallart}, {Zoccali}, \&
  {Aparicio}}]{Gallart2005}
{Gallart}, C., {Zoccali}, M., \& {Aparicio}, A. 2005, \araa, 43, 387

\bibitem[{{Gnedin}(2003)}]{Gnedin2003}
{Gnedin}, O.~Y. 2003, \apj, 589, 752

\bibitem[{{Guhathakurta} \& {Tyson}(1989)}]{Guhathakurta1989}
{Guhathakurta}, P., \& {Tyson}, J.~A. 1989, \apj, 346, 773

\bibitem[{{Gunn} \& {Gott}(1972)}]{GunnGott1972}
{Gunn}, J.~E., \& {Gott}, III, J.~R. 1972, \apj, 176, 1

\bibitem[{{Heymans} {et~al.}(2012){Heymans}, {Van Waerbeke}, {Miller}, {Erben},
  {Hildebrandt}, {Hoekstra}, {Kitching}, {Mellier}, {Simon}, {Bonnett},
  {Coupon}, {Fu}, {Harnois D{\'e}raps}, {Hudson}, {Kilbinger}, {Kuijken},
  {Rowe}, {Schrabback}, {Semboloni}, {van Uitert}, {Vafaei}, \&
  {Velander}}]{Heymans2012}
{Heymans}, C., {Van Waerbeke}, L., {Miller}, L., {et~al.} 2012, \mnras, 427,
  146

\bibitem[{{Impey} {et~al.}(1988){Impey}, {Bothun}, \& {Malin}}]{Impey1988}
{Impey}, C., {Bothun}, G., \& {Malin}, D. 1988, \apj, 330, 634

\bibitem[{{Impey} {et~al.}(1996){Impey}, {Sprayberry}, {Irwin}, \&
  {Bothun}}]{Impey1996}
{Impey}, C.~D., {Sprayberry}, D., {Irwin}, M.~J., \& {Bothun}, G.~D. 1996,
  \apjs, 105, 209

\bibitem[{{Javanmardi} {et~al.}(2016){Javanmardi}, {Martinez-Delgado},
  {Kroupa}, {Henkel}, {Crawford}, {Teuwen}, {Gabany}, {Hanson}, {Chonis}, \&
  {Neyer}}]{Javanmardi2016}
{Javanmardi}, B., {Martinez-Delgado}, D., {Kroupa}, P., {et~al.} 2016, \aap,
  588, A89

\bibitem[{{Karachentsev} {et~al.}(2015){Karachentsev}, {Riepe}, {Zilch},
  {Blauensteiner}, {Elvov}, {Hochleitner}, {Hubl}, {Kerschhuber},
  {K{\"u}ppers}, {Neyer}, {P{\"o}lzl}, {Remmel}, {Schneider}, {Sparenberg},
  {Trulson}, {Willems}, \& {Ziegler}}]{Karachentsev2015}
{Karachentsev}, I.~D., {Riepe}, P., {Zilch}, T., {et~al.} 2015, Astrophysical
  Bulletin, 70, 379

\bibitem[{{Koda} {et~al.}(2015){Koda}, {Yagi}, {Yamanoi}, \&
  {Komiyama}}]{Koda2015}
{Koda}, J., {Yagi}, M., {Yamanoi}, H., \& {Komiyama}, Y. 2015, \apjl, 807, L2

\bibitem[{{Magnier} \& {Cuillandre}(2004)}]{Magnier2004}
{Magnier}, E.~A., \& {Cuillandre}, J.-C. 2004, \pasp, 116, 449

\bibitem[{{Makarov} \& {Karachentsev}(2011)}]{MK2011}
{Makarov}, D., \& {Karachentsev}, I. 2011, \mnras, 412, 2498

\bibitem[{{Makarov} {et~al.}(2006){Makarov}, {Makarova}, {Rizzi}, {Tully},
  {Dolphin}, {Sakai}, \& {Shaya}}]{Makarov2006}
{Makarov}, D., {Makarova}, L., {Rizzi}, L., {et~al.} 2006, \aj, 132, 2729

\bibitem[{{Makarov} {et~al.}(2014){Makarov}, {Prugniel}, {Terekhova},
  {Courtois}, \& {Vauglin}}]{LEDA}
{Makarov}, D., {Prugniel}, P., {Terekhova}, N., {Courtois}, H., \& {Vauglin},
  I. 2014, \aap, 570, A13

\bibitem[{{Makarov} {et~al.}(2015){Makarov}, {Sharina}, {Karachentseva}, \&
  {Karachentsev}}]{Makarov2015}
{Makarov}, D.~I., {Sharina}, M.~E., {Karachentseva}, V.~E., \& {Karachentsev},
  I.~D. 2015, \aap, 581, A82

\bibitem[{{Mart{\'{\i}}nez-Delgado} {et~al.}(2016){Mart{\'{\i}}nez-Delgado},
  {L{\"a}sker}, {Sharina}, {Toloba}, {Fliri}, {Beaton}, {Valls-Gabaud},
  {Karachentsev}, {Chonis}, {Grebel}, {Forbes}, {Romanowsky},
  {Gallego-Laborda}, {Teuwen}, {G{\'o}mez-Flechoso}, {Wang}, {Guhathakurta},
  {Kaisin}, \& {Ho}}]{MartinezDelgado2016}
{Mart{\'{\i}}nez-Delgado}, D., {L{\"a}sker}, R., {Sharina}, M., {et~al.} 2016,
  \aj, 151, 96

\bibitem[{{Mateo}(1998)}]{Mateo1998}
{Mateo}, M.~L. 1998, \araa, 36, 435

\bibitem[{{McConnachie}(2012)}]{McConnachie2012}
{McConnachie}, A.~W. 2012, \aj, 144, 4

\bibitem[{{McConnachie} {et~al.}(2008){McConnachie}, {Huxor}, {Martin},
  {Irwin}, {Chapman}, {Fahlman}, {Ferguson}, {Ibata}, {Lewis}, {Richer}, \&
  {Tanvir}}]{McConnachie2008}
{McConnachie}, A.~W., {Huxor}, A., {Martin}, N.~F., {et~al.} 2008, \apj, 688,
  1009

\bibitem[{{McGaugh} \& {Bothun}(1994)}]{McGaugh1994}
{McGaugh}, S.~S., \& {Bothun}, G.~D. 1994, \aj, 107, 530

\bibitem[{{Merritt} {et~al.}(2014){Merritt}, {van Dokkum}, \&
  {Abraham}}]{Merritt2014}
{Merritt}, A., {van Dokkum}, P., \& {Abraham}, R. 2014, \apjl, 787, L37

\bibitem[{{Merritt} {et~al.}(2016){Merritt}, {van Dokkum}, {Abraham}, \&
  {Zhang}}]{Merritt2016}
{Merritt}, A., {van Dokkum}, P., {Abraham}, R., \& {Zhang}, J. 2016, ArXiv:1606.08847

\bibitem[{{Mihos} {et~al.}(2015){Mihos}, {Durrell}, {Ferrarese}, {Feldmeier},
  {C{\^o}t{\'e}}, {Peng}, {Harding}, {Liu}, {Gwyn}, \&
  {Cuillandre}}]{Mihos2015}
{Mihos}, J.~C., {Durrell}, P.~R., {Ferrarese}, L., {et~al.} 2015, \apjl, 809,
  L21

\bibitem[{{Moore} {et~al.}(1996){Moore}, {Katz}, {Lake}, {Dressler}, \&
  {Oemler}}]{Moore1996}
{Moore}, B., {Katz}, N., {Lake}, G., {Dressler}, A., \& {Oemler}, A. 1996,
  \nat, 379, 613

\bibitem[{{Moore} {et~al.}(1999){Moore}, {Lake}, {Stadel}, \&
  {Quinn}}]{Moore1999}
{Moore}, B., {Lake}, G., {Stadel}, J., \& {Quinn}, T. 1999, in Astronomical
  Society of the Pacific Conference Series, Vol. 170, The Low Surface
  Brightness Universe, ed. J.~I. {Davies}, C.~{Impey}, \& S.~{Phillips}, 229

\bibitem[{{Mu{\~n}oz} {et~al.}(2015){Mu{\~n}oz}, {Eigenthaler}, {Puzia},
  {Taylor}, {Ordenes-Brice{\~n}o}, {Alamo-Mart{\'{\i}}nez}, {Ribbeck},
  {{\'A}ngel}, {Capaccioli}, {C{\^o}t{\'e}}, {Ferrarese}, {Galaz}, {Hempel},
  {Hilker}, {Jord{\'a}n}, {Lan{\c c}on}, {Mieske}, {Paolillo}, {Richtler},
  {S{\'a}nchez-Janssen}, \& {Zhang}}]{Munoz2015}
{Mu{\~n}oz}, R.~P., {Eigenthaler}, P., {Puzia}, T.~H., {et~al.} 2015, \apjl,
  813, L15

\bibitem[{{O'Neil} {et~al.}(1997{\natexlab{a}}){O'Neil}, {Bothun}, \&
  {Cornell}}]{ONeil1997a}
{O'Neil}, K., {Bothun}, G.~D., \& {Cornell}, M.~E. 1997{\natexlab{a}}, \aj,
  113, 1212

\bibitem[{{O'Neil} {et~al.}(1997{\natexlab{b}}){O'Neil}, {Bothun}, {Schombert},
  {Cornell}, \& {Impey}}]{ONeil1997b}
{O'Neil}, K., {Bothun}, G.~D., {Schombert}, J., {Cornell}, M.~E., \& {Impey},
  C.~D. 1997{\natexlab{b}}, \aj, 114, 2448

\bibitem[{{Peng} \& {Lim}(2016)}]{PengLim2016}
{Peng}, E.~W., \& {Lim}, S. 2016, \apjl, 822, L31

\bibitem[{{Roman} \& {Trujillo}(2016)}]{RomanTrujillo2016}
{Roman}, J., \& {Trujillo}, I. 2016, ArXiv:1603.03494

\bibitem[{{Shappee} \& {Stanek}(2011)}]{Shappee2011}
{Shappee}, B.~J., \& {Stanek}, K.~Z. 2011, The Astrophysical Journal, 733, 124

\bibitem[{{Toloba} {et~al.}(2016){Toloba}, {Sand}, {Spekkens}, {Crnojevi{\'c}},
  {Simon}, {Guhathakurta}, {Strader}, {Caldwell}, {McLeod}, \&
  {Seth}}]{Toloba2016}
{Toloba}, E., {Sand}, D.~J., {Spekkens}, K., {et~al.} 2016, \apjl, 816, L5

\bibitem[{{Tully}(2015)}]{Tully2015}
{Tully}, R.~B. 2015, \aj, 149, 171

\bibitem[{{Tully} {et~al.}(2016){Tully}, {Courtois}, \& {Sorce}}]{Tully2016}
{Tully}, R.~B., {Courtois}, H.~M., \& {Sorce}, J.~G. 2016, \aj, 152, 50

\bibitem[{{Ulmer} {et~al.}(1996){Ulmer}, {Bernstein}, {Martin}, {Nichol},
  {Pendleton}, \& {Tyson}}]{Ulmer1996}
{Ulmer}, M.~P., {Bernstein}, G.~M., {Martin}, D.~R., {et~al.} 1996, \aj, 112,
  2517

\bibitem[{{van den Hoek} {et~al.}(2000){van den Hoek}, {de Blok}, {van der
  Hulst}, \& {de Jong}}]{vandenHoek2000}
{van den Hoek}, L.~B., {de Blok}, W.~J.~G., {van der Hulst}, J.~M., \& {de
  Jong}, T. 2000, \aap, 357, 397

\bibitem[{{van der Burg} {et~al.}(2016){van der Burg}, {Muzzin}, \&
  {Hoekstra}}]{vanderBurg2016}
{van der Burg}, R.~F.~J., {Muzzin}, A., \& {Hoekstra}, H. 2016, \aap, 590, A20

\bibitem[{{van Dokkum} {et~al.}(2016){van Dokkum}, {Abraham}, {Brodie},
  {Conroy}, {Danieli}, {Merritt}, {Mowla}, {Romanowsky}, \&
  {Zhang}}]{vanDokkum2016}
{van Dokkum}, P., {Abraham}, R., {Brodie}, J., {et~al.} 2016, \apjl, 828, L6

\bibitem[{{van Dokkum} {et~al.}(2014){van Dokkum}, {Abraham}, \&
  {Merritt}}]{vanDokkum2014}
{van Dokkum}, P.~G., {Abraham}, R., \& {Merritt}, A. 2014, \apjl, 782, L24

\bibitem[{{van Dokkum} {et~al.}(2015){van Dokkum}, {Abraham}, {Merritt},
  {Zhang}, {Geha}, \& {Conroy}}]{vanDokkum2015a}
{van Dokkum}, P.~G., {Abraham}, R., {Merritt}, A., {et~al.} 2015, \apjl, 798,
  L45

\bibitem[{{Vijayaraghavan} \& {Ricker}(2013)}]{Vijayaraghavan2013}
{Vijayaraghavan}, R., \& {Ricker}, P.~M. 2013, \mnras, 435, 2713

\bibitem[{{Villegas} {et~al.}(2008){Villegas}, {Kissler-Patig}, {Jord{\'a}n},
  {Goudfrooij}, \& {Zwaan}}]{Villegas2008}
{Villegas}, D., {Kissler-Patig}, M., {Jord{\'a}n}, A., {Goudfrooij}, P., \&
  {Zwaan}, M. 2008, \aj, 135, 467

\bibitem[{{Wetzel} {et~al.}(2015){Wetzel}, {Deason}, \&
  {Garrison-Kimmel}}]{Wetzel2015}
{Wetzel}, A.~R., {Deason}, A.~J., \& {Garrison-Kimmel}, S. 2015, \apj, 807, 49

\bibitem[{{Yagi} {et~al.}(2016){Yagi}, {Koda}, {Komiyama}, \&
  {Yamanoi}}]{Yagi2016}
{Yagi}, M., {Koda}, J., {Komiyama}, Y., \& {Yamanoi}, H. 2016, \apjs, 225, 11

\bibitem[{{Zucker} {et~al.}(2006){Zucker}, {Belokurov}, {Evans}, {Kleyna},
  {Irwin}, {Wilkinson}, {Fellhauer}, {Bramich}, {Gilmore}, {Newberg}, {Yanny},
  {Smith}, {Hewett}, {Bell}, {Rix}, {Gnedin}, {Vidrih}, {Wyse}, {Willman},
  {Grebel}, {Schneider}, {Beers}, {Kniazev}, {Barentine}, {Brewington},
  {Brinkmann}, {Harvanek}, {Kleinman}, {Krzesinski}, {Long}, {Nitta}, \&
  {Snedden}}]{Zucker2006}
{Zucker}, D.~B., {Belokurov}, V., {Evans}, N.~W., {et~al.} 2006, \apjl, 650,
  L41

\end{thebibliography}

\end{document}